\documentclass[11pt,a4paper]{article}

\usepackage{amsmath,amssymb,epsfig,cite}

\def\lp{\left(}
\def\rp{\right)}
\def\lb{\left[}
\def\rb{\right]}
\def\be{\begin{equation}}
\def\ee{\end{equation}}
\begin{document}

\title{Stability of charged thin shells} 
\author{Ernesto F. Eiroa$^{1,2,}$\thanks{e-mail: eiroa@iafe.uba.ar}, 
Claudio Simeone$^{2,3,}$\thanks{e-mail: csimeone@df.uba.ar}\\
{\small $^1$ Instituto de Astronom\'{\i}a y F\'{\i}sica del Espacio, C.C. 67, Suc. 28, 1428, Buenos Aires, Argentina}\\
{\small $^2$ Departamento de F\'{\i}sica, Facultad de Ciencias Exactas y 
Naturales,} \\ 
{\small Universidad de Buenos Aires, Ciudad Universitaria Pab. I, 1428, 
Buenos Aires, Argentina} \\
{\small $^3$ IFIBA, CONICET, Ciudad Universitaria Pab. I, 1428, 
Buenos Aires, Argentina}} 
\maketitle

\begin{abstract}

In this article we study the mechanical stability of spherically symmetric thin shells with charge, in Einstein--Maxwell and Einstein--Born--Infeld theories. We analyze linearized perturbations preserving the symmetry, for shells around vacuum and shells surrounding non charged black holes. \\

\noindent 
PACS number(s): 04.20.Gz; 04.40.-b; 04.70.Bw 

\end{abstract}

\section{Introduction}

Since the leading works by Darmois and Israel \cite{daris} introducing the formalism for surface layers, the study of thin-shell characterization and dynamics in general relativity have received increasing attention. The Lanczos equations \cite{daris,mus} which relate the surface energy-momentum tensor of a shell with the geometry at both sides of it --or, more precisely, with the jump of the extrinsic curvature tensor across it-- have been widely applied within different frameworks. Apart from cosmological applications, the Darmois--Israel formalism has become the central tool for the study of the dynamics and matter characterization of highly symmetric configurations. The linearized stability analysis of spherical shells  was carried out by several authors (see  \cite{poi,isla,gon,locr} and the references included in these works).  The formalism was applied to bubbles, shells around stars and black holes, and in the construction of thin-shell wormholes (with spherical, plane and also cylindrical throats; see for example \cite{book,wh1,wh2,wh3,wh4,dile,lelo,whc} and references therein). Besides, thin shells are associated to gravastars, for which the stability was also studied \cite{gravastars}.

In order to avoid the infinite self energies of charged point particles in Maxwell description of the electromagnetic field, Born and Infeld introduced in 1934 a nonlinear theory of electrodynamics \cite{borninf}. The corresponding spherically symmetric solution for the metric of a charged object was first obtained in 1935 by Hoffmann \cite{hoffmann}. It was shown, however, that Hoffmann solution does not provide a suitable classical model for the electron, but it corresponds to a black hole.  The interest in nonlinear electrodynamics was renewed because Born--Infeld type actions appeared in low energy string theory \cite{nle}. Spherically symmetric black holes in Born--Infeld electrodynamics coupled to Einstein gravity were studied in recent years by several authors \cite{gibbons,bibh,breton}. Also, the stability of thin-shell wormholes has been recently considered \cite{risi09}. 

In this work, we address the mechanical stability of spherical shells under perturbations preserving the symmetry within the frameworks of Einstein--Maxwell and Einstein--Born--Infeld theories. We first adapt the formalism to the form most suitable for our purposes. Then we perform the mathematical construction of shells starting from the Reissner--Nordstr\"{o}m solution and from the corresponding extension of this metric for Born--Infeld electrodynamics. Finally, we analyze in detail the cases of charged bubbles, and charged layers around non charged black holes. In what follows we adopt units such that $c=G=1$.

\section{Spherical shells: formalism}

For the study of spherically symmetric shells we follow the standard approach \cite{daris,mus,poi,isla,gon,locr,book,wh1} in which the matter layer appears as a result of cutting and pasting two manifolds $\mathcal{M}_1$ and $\mathcal{M}_2$ to construct a new geodesically complete one $\mathcal{M}$. We start from the metrics
\begin{equation}
ds_{1,2}^2 = -f_{1,2}(r_{1,2})dt_{1,2}^2 +f_{1,2}^{-1}(r_{1,2})dr_{1,2}^2+r_{1,2}^2(d\theta^2+\sin^2\theta d\varphi^2),
\label{e1}
\end{equation}
and we paste them at the spherical surface $\Sigma$ defined by $r_{1,2}=a$. The manifold ${\cal M}$ is given by the union of the inner part ($r_1\leq a$) of $\mathcal{M}_1$ and by the outer part ($r_2\geq a$) of $\mathcal{M}_2$ . The line element is continuous across $\Sigma$ as the coordinates in each side are chosen to satisfy $f_1(a)dt_1^2=f_2(a)dt_2^2$. For the study of the stability of our construction, we let the radius $a$ to be a function of the proper time $\tau$ measured by  an observer on the surface. The induced metric on $\Sigma$ is unique, and it has the form
\be
ds_\Sigma^2=-d\tau^2+a^2(\tau)(d\theta ^2+\sin^2\theta d\varphi ^2).
\ee
The coordinates of the embedding are $X_{1,2}^\mu=(t_{1,2},r_{1,2},\theta,\varphi)$ while the coordinates at the surface $\Sigma$ are $\xi ^i=(\tau , \theta,\varphi )$. The relation between the geometry and the matter on the surface is given by the Lanczos equations \cite{daris,mus}  
\begin{equation}
-[K_{ij}]+g_{ij}[K]=8\pi S_{ij}
\label{e7}
\end{equation}
where $g_{ij}$ is the induced metric on $\Sigma$, $K_{ij}$ is the extrinsic curvature, $K$ is its trace, and $S_{ij}$ is the surface energy-momentum tensor; the brackets denote the jump of a given quantity $q$ across the surface: $[q]=q^2\vert_\Sigma -q^1\vert_\Sigma$. It is clear that $[g_{ij}]=0 $, i.e. the geometry is continuous across $\Sigma$ as required by the thin-shell formalism \cite{mus}. If $[K_{ij}]=0$ we speak of $\Sigma$ as a boundary surface, while if $[K_{ij}]\neq 0$ we have a thin shell at $\Sigma$, where a layer of matter is present.  The general form of the components of  $K_{ij}$ at each side of the shell is given by
\begin{equation}
K_{ij}^{1,2} = - n_{\gamma}^{1,2} \left. \left( \frac{\partial^2
X^{\gamma}_{1,2}}{\partial \xi^i \partial \xi^j} +\Gamma_{\alpha\beta}^{\gamma}
\frac{\partial X^{\alpha}_{1,2}}{\partial \xi^i} \frac{\partial
X^{\beta}_{1,2}}{\partial \xi^j} \right) \right|_{\Sigma}, \label{e2}
\end{equation}
where $n_{\gamma}^{1,2}$ are unit normals ($n^{\gamma} n_{\gamma} = 1$) to the surface. If we define ${\cal H}(r_{1,2})=r_{1,2}-a(\tau)=0$, we have
\begin{equation}
n_{\gamma}^{1,2} = \left| g^{\alpha \beta} \frac{\partial
\mathcal{H}}{\partial X^{\alpha}_{1,2}} \frac{\partial \mathcal{H}}{\partial
X^{\beta}_{1,2}} \right|^{- 1 / 2} \frac{\partial \mathcal{H}}{\partial
X^{\gamma}_{1,2}}, \label{e3}
\end{equation}
where the unit normals at both sides of $\Sigma $ are oriented outwards from the origin.  The normal to $\Sigma$ is unique and points from region 1 to region 2 as required by the sign convention adopted in Eq. (\ref{e7}). For the particular form (\ref{e1}) of the metric, Eqs. (\ref{e7}), (\ref{e2}) and (\ref{e3}) give the following  components of the surface energy momentum tensor:
\be
S_\tau^\tau=-\sigma= \frac{1}{4\pi a}\lp\sqrt{f_2(a)+{\dot a}^2}-\sqrt{f_1(a)+{\dot a}^2}\rp.\label{00}
\ee
\be
S_\theta^\theta=S_\varphi^\varphi=p=-\frac{\sigma}{2}+\frac{1}{16\pi}\lp\frac{2\ddot a+f'_2(a)}{\sqrt{f_2(a)+{\dot a}^2}}-\frac{2\ddot a+f'_1(a)}{\sqrt{f_1(a)+{\dot a}^2}}\rp.\label{pres}
\ee
We have adopted the usual notation in which  the prime represents  $d/dr$, and the dot stands for $d/d\tau $. The equations above, or any of them plus the conservation equation
\be
\frac{d(a^2\sigma)}{d\tau}+p\frac{da^2}{d\tau}=0,
\label{cons}
\ee
which is valid because of the form of the metric\footnote{in other terms, the shell is transparent \cite{isla}.}, determine the evolution of the shell radius as a function of the proper time. They are the starting point for the analysis of the mechanical stability of thin shells. We now consider small perturbations preserving the symmetry around a static solution of the equations above. Our procedure is similar to the treatment in Refs. \cite{poi,isla,gon,locr,wh1}\footnote{In Ref. \cite{isla}, Eqs. (17) and (18) (and the results derived from them) are not correct, because $p'/\sigma '$ is present in both sides of Eq. (15) through the second derivative of the surface mass of the shell, but it only appears in one side of Eqs. (17) and (18). The same applies to Eq. (47) of Ref. \cite{locr}. The correct result is shown (using a different notation) in Eq. (5.19) of Ref. \cite{gon}.}. Provided the equation of state $p=p(\sigma)$, the conservation equation can be formally integrated \cite{wh1} to give $\sigma=\sigma(a)$.  After some algebraic manipulations, from Eq. (\ref{00}) we then obtain
\be
{\dot a}^2+V(a)=0,\label{energy}
\ee
where
\be
V(a)=\frac{f_1(a)+f_2(a)}{2}-\lp2\pi a \sigma\rp^2-\lp\frac{f_1(a)-f_2(a)}{8\pi a \sigma}\rp^2
\ee
is commonly understood as a potential, given the analogy between Eq. (\ref{energy}) and the energy of a point particle with only one degree of freedom. Defining $S(a)=\lp f_1(a)+f_2(a)\rp /2$, $R(a)=\lp f_1(a)-f_2(a)\rp /2$ and $m(a)=4\pi a^2\sigma$, 
the potential has the form
\be
V(a)=S-\frac{1}{4}\lp\frac{m}{a}\rp^2-\lp\frac{a}{m}\rp^2 R^2.
\ee
For a perturbative treatment of the stability of static solutions it is enough with the analysis of the first and second derivatives of the potential at a radius $a_0$ for which $\dot a=0$. Equilibrium satisfies $V(a_0)=0$ and $V'(a_0)=0$, and stability requires $V''(a_0)>0$. After evaluating the derivatives and some algebraic manipulations, the condition for stability gives 
\be
\frac{m}{4a_0}\lp\frac{m}{a_0}\rp''+ \frac{a_0}{m}\lp\frac{a_0}{m}\rp''R^2 < \Omega(a_0)-\Gamma^2(a_0),
 \ee 
where
\be
\Gamma(a_0) = \frac{a_0}{m}\lb S'-2\frac{a_0}{m}\lp\frac{a_0}{m}\rp'R^2-2\lp\frac{a_0}{m}\rp^2 RR'\rb
\ee
and
\be
\Omega(a_0)  =  \frac{S''}{2}-\lb\lp\frac{a_0}{m}\rp'\rb^2R^2 -4\frac{a_0}{m}\lp\frac{a_0}{m}\rp'RR' - \lp\frac{a_0}{m}\rp^2\lb R'^2+RR''\rb.
\ee
In these expressions $m$, $S$ and $R$ are given as functions of $a_0$ and the primes stand for derivatives with respect to $a_0$. From the conservation equation (\ref{cons}) we have $\sigma '=-2(\sigma + p)/a_0$, then $(m/a_0)'=-4\pi(\sigma+2p)$, so we obtain
\be
\lp\frac{m}{a_0}\rp''=-4\pi \sigma'(a_0) \lp 1+2\eta \rp,
\ee
and
\be
\lp\frac{a_0}{m}\rp''= 2 \lp \frac{a_0}{m}\rp ^3 \lb \lp \frac{m}{a_0} \rp ' \rb ^2 + 4 \pi \lp \frac{a_0}{m}\rp ^2 \sigma'(a_0) \lp 1+2\eta \rp,
\ee
where $\eta = p'(a_0)/\sigma'(a_0)$. From the definition of $m$ we have
\be
\sigma'(a_0)=\frac{1}{4\pi a_0}\lb\lp\frac{m}{a_0}\rp'-\frac{m}{a_0^2}\rb,
\ee
so that the condition for stable equilibrium can be put in the form
\be
\chi (a_0)<\Omega(a_0)-\Gamma^2(a_0),
\label{sta}
\ee
where
\be
\chi (a_0)=2 \lp \frac{a_0}{m} \rp ^4 \lb \lp\frac{m}{a_0}\rp' \rb ^2 R^2 + \frac{1}{a_0}\lb \frac{m}{4a_0} - \lp \frac{a_0}{m}\rp ^3 R^2 \rb \lb \frac{m}{a_0^2}-\lp\frac{m}{a_0}\rp'\rb \lp 1+2\eta \rp.
\ee 
In what follows, the analysis is carried out in terms of the parameter $\eta $, which in the case that $0<\eta\leq 1$ can be interpreted as the square of the velocity of sound on the shell.

The construction and the stability analysis presented above are also valid for wormholes as long as the outer part of both manifolds is taken, and the minus sign inside the parenthesis in Eqs. (\ref{00}) and (\ref{pres}) is replaced by a plus sign. In particular, if $\mathcal{M}_{1,2}$ are equal copies of the geometry defined by Eq. (\ref{e1}) with $r\geq a$ the results of Refs. \cite{wh1,wh2,wh3,wh4,risi09} can be recovered.

\section{Charged shells: construction and stability}

We now study thin shells in Einstein--Maxwell and Einstein--Born--Infeld theories.  The field equations of Einstein gravity coupled to Born--Infeld nonlinear electrodynamics have the vacuum spherically symmetric solution \cite{gibbons,breton} given by the metric
\begin{equation}
ds^{2}=-f(r)dt^{2}+f^{-1}(r)dr^{2}+r^{2}(d\theta^2+\sin^2\theta d\varphi^2),
\label{bi1}
\end{equation}
with
\begin{equation}
f(r)=1-\frac{2M}{r}+\frac{2}{3b^{2}}\left\{ r^{2}-\sqrt{r^{4}+b^{2}Q^{2}}+
\frac{\sqrt{|bQ|^{3}}}{r}F\left[ \arccos\left( \frac{r^{2}-|bQ|}{r^{2}+|bQ|}
\right) ,\frac{\sqrt{2}}{2}\right]\right\} ,
\label{bi2}
\end{equation} 
where $F(\gamma ,k)$ is the elliptic integral of the first kind\footnote{
$F(\gamma , k)=\int _{0}^{\gamma }(1-k^{2}\sin ^{2}\phi )^{-1/2}d\phi =
\int _{0}^{\sin \gamma }[(1-z^{2})(1-k^{2}z^{2})]^{-1/2}dz $ 
(See Ref. \cite{gradshteyn}).}, and the electric and magnetic inductions 
\begin{equation}
D(r)=\frac{Q_{E}}{r^{2}},
\label{bi2a}
\end{equation}
\begin{equation}
B(r)=Q_{M}\sin \theta.
\label{bi2b}
\end{equation}
As usual, $M>0$ stands for the ADM mass, and $Q^{2}=Q_{E}^{2}+Q_{M}^{2}$ is the sum of the squares of the electric $Q_{E}$ and magnetic $Q_{M}$ charges. With the units adopted above, $M$, $Q$ and $b$ have dimensions of length. The parameter $b$ indicates how much Born--Infeld electrodynamics differs from Maxwell theory. In the limit $b\rightarrow 0$, the Reissner--Nordstr\"{o}m metric is obtained:
\be
f(r)=1-\frac{2M}{r}+\frac{Q^2}{r^2}.
\label{rn}
\ee
The metric given by Eqs. (\ref{bi1}) and (\ref{bi2}) is also asymptotically Reissner--Nordstr\"{o}m for large values of $r$. The geometry is singular at the origin \cite{breton}, as it happens with the Schwarzschild and Reissner--Nordstr\"{o}m cases. The zeros of the function $f(r)$ correspond to the horizons, which must be found numerically. For a given value of the constant $b$, when the charge is small, $0\le |Q|/M\le \nu _{1}$, the function $f(r)$ has only one zero and there is a regular event horizon. For intermediate values of charge, $\nu _{1}<|Q|/M<\nu _{2}$, $f(r)$ has two zeros; then, as it happens for the Reissner--Nordstr\"{o}m geometry, an inner horizon and an outer regular event horizon exist. When $|Q|/M=\nu _{2}$, there is one degenerate horizon. Finally, if the values of charge are large, $|Q|/M>\nu _{2}$, the function $f(r)$ has no zeros and a naked singularity appears. The values of $|Q|/M$ where the number of horizons change, $\nu _{1}=(9|b|/M)^{1/3} [F(\pi ,\sqrt{2}/2]^{-2/3}$ and $\nu _{2}$, which must be obtained numerically from the condition $f(r_{h})=f'(r_{h})=0$, are increasing functions of $|b|/M$. In the Reissner--Nordstr\"{o}m limit ($b\rightarrow 0$) it is easy to see that $\nu _{1}=0$ and $\nu _{2}=1$.

\begin{figure}[t!]
\centering
\includegraphics[width=16.5cm]{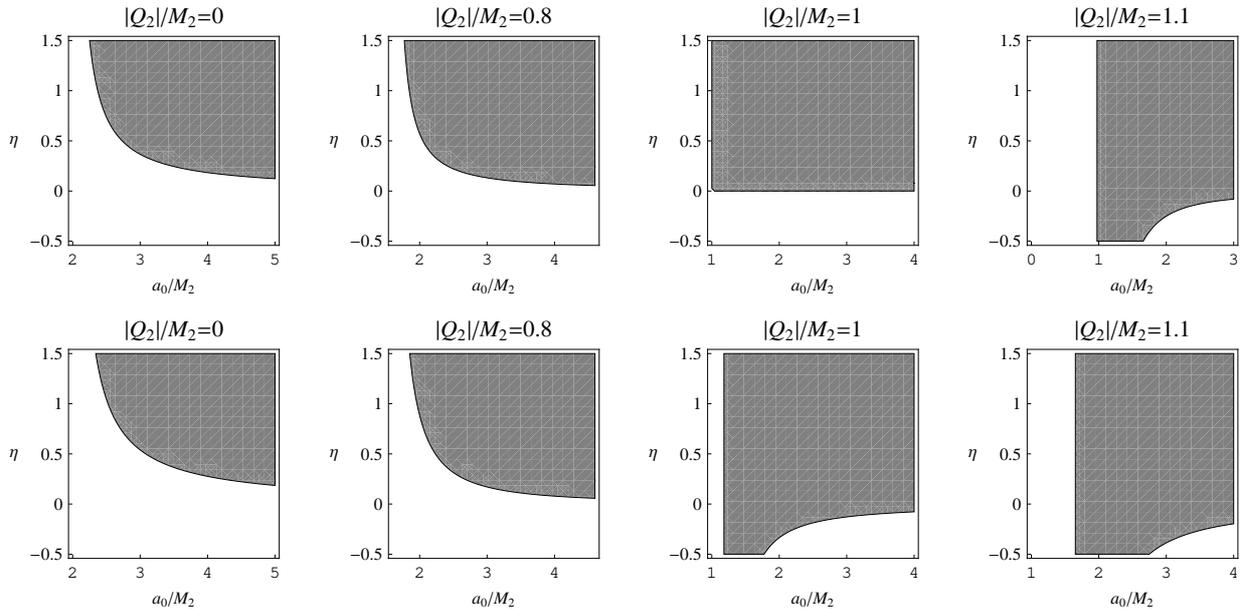}
\caption{Stability regions (grey) for charged shells satisfying the weak energy condition in Einstein--Maxwell theory, around vacuum  ($M_1=0$, $Q_1=0$)  shown in the upper row figures, and around a black hole ($M_1/M_2=0.5$, $Q_1=0$)  shown in the lower row figures.}
\label{frn}
\end{figure}

We start our construction from Eq. (\ref{bi1}), with the metric function given by Eq. (\ref{bi2}) or Eq.  (\ref{rn}); as it was stated above, the geometry consists in the inner part of manifold $\mathcal{M}_1$ and the outer part of manifold $\mathcal{M}_2$. In all cases we restrict our analysis to normal matter, so that  the weak energy  condition, i.e. $\sigma\geq 0$ and $\sigma+p\geq 0$, is fulfilled. We first consider shells around vacuum (bubbles), in which we take the outer manifold with mass $M_2$ and charge $Q_2$, and the inner one with mass $M_1=0$ and charge $Q_1=0$. The radius $a_0$ is chosen larger than the horizon radius of the outer manifold (so that the singularity and the event horizon of the original manifold are both removed). The second kind of configuration that we analyze are charged shells around non charged black holes, so the inner geometry has mass $M_1$ and no charge, and the outer one has mass $M_2$ and charge $Q_2$. In this case, besides demanding that $a_0$ is greater than the horizon radius of the outer manifold, we also have to take $a_0>2M_1$. A necessary condition (but not sufficient in the case of charged shells) for fulfilling the weak energy condition is that $0\leq M_1<M_2$. As pointed out above, when the parameter $\eta $ is within the range  $0<\eta \le 1$, it can be interpreted as the square of the  velocity of sound on the shell. If $\eta >1$ this interpretation is not valid, because it would mean a speed greater than the velocity of light, implying the violation of causality. Matter with $\eta <0$ is not common (though not impossible, e.g. $\eta =-1$ in the Casimir effect between the plates). In such a case, clearly the interpretation of $\eta $ as the squared velocity of sound is no longer admissible. In what follows we consider any value of $\eta $, but we are more interested in the results corresponding to $0<\eta \le 1$.

\begin{figure}[t!]
\centering
\includegraphics[width=16.5cm]{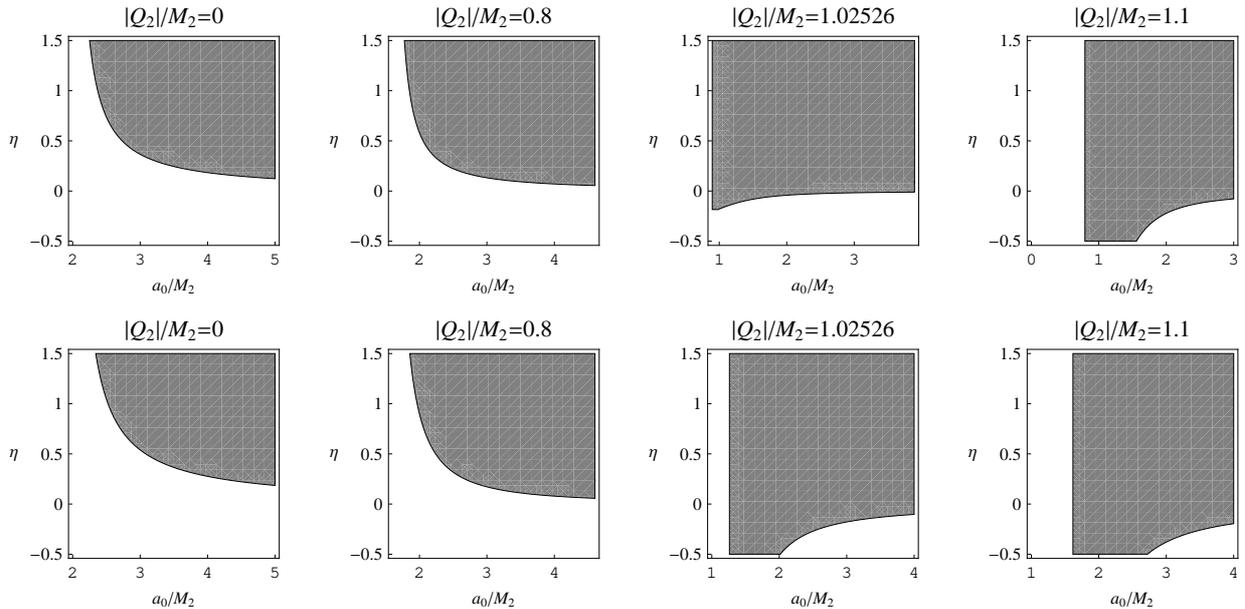}
\caption{Stability regions (grey) for charged shells satisfying the weak energy condition in Einstein--Born--Infeld  theory with  $b=1$, around vacuum  ($M_1=0$, $Q_1=0$)  shown in the upper row figures, and around a black hole ($M_1/M_2=0.5$, $Q_1=0$)  shown in the lower row figures.}
\label{fbi}
\end{figure}

The interpretation of the stability results from Eq. (\ref{sta}) is not straightforward, so we present them graphically. In Einstein--Maxwell theory the stability regions are shown in Fig. \ref{frn}, while the results within the framework of Einstein--Born--Infeld theory are shown in Fig. \ref{fbi}; we consider a large deviation from Maxwell electrodynamics by setting the Born--Infeld constant $b=1$. The figures correspond to some representative cases which illustrate the dependence of the stability regions --for matter satisfying the weak energy condition-- in terms of the parameters and the constant $b$; in particular, for the case of shells surrounding black holes we show the results corresponding to $M_1/M_2=0.5$. For comparison, the non charged shells results previously studied in Ref. \cite{poi} are also displayed in the figures. In both theories, the qualitative behaviour is different if the charge is under or beyond the critical value (from which the horizon of the outer original manifold vanishes). If the charge is under the critical value, stable configurations are possible only for positive $\eta $, while if the charge is equal or above the critical value, negative values of $\eta $ are compatible with stability. In all cases, there are values of $a_0/M_2$ for which stable configurations are possible with $0<\eta \leq 1$.  Within this range of $\eta $, if the charge is below the critical value the largest interval of $a_0/M_2$ for which the shell is stable corresponds to $\eta =1$, as it was obtained for non charged shells in Ref. \cite{poi}. In both electrodynamics, when the charge is under the critical value, shells around black holes present slightly smaller regions of stability than bubbles; and the stability regions become larger as the charge increases\footnote{The same happens for spherical thin-shell wormholes with charge \cite{wh2}.}. If the charge is equal or beyond the critical value, for a given $\eta$ bubbles can be stable for smaller radii $a_0/M_2$ than shells around black holes, and for fixed $a_0/M_2$ bubbles are stable for a smaller range of the parameter $\eta $ than shells around black holes. The stability regions corresponding to bubbles and shells around black holes have similar forms in Maxwell and Born--Infeld electrodynamics. The main difference between the two theories is the value of the critical charge (which is a growing function of $b$) where the form of the stability regions change, as pointed out above.

\section{Summary} 

We have studied the stability of charged spherical shells within the frameworks of Einstein gravity coupled to Maxwell and to Born--Infeld electrodynamic theories. The starting point has been the mathematical construction of a shell from two spherically symmetric geometries associated to different masses and charges. For both bubbles (shells around vacuum) and shells around non charged black holes we have considered linearized perturbations preserving the  symmetry. The analysis has been carried out by drawing the stability regions in terms of the shell radius $a_0$ and  $\eta=p'(a_0)/\sigma'(a_0)$, for different values of mass and charge. In both theories, the presence of the charge seems to enlarge the stability regions  for both bubbles and shells around black holes. The stability regions for bubbles appear to be larger than those of shells around black holes if the charge is under the critical value, while the reverse is true for charges  above the critical value. According to the results displayed, shells in both electromagnetic theories show a similar behavior. We have shown that charged layers with $0<\eta\leq 1$ can be stable for suitable values of the parameters.

\section*{Acknowledgments}

This work has been supported by Universidad de Buenos Aires and CONICET.

\end{document}